%
%
%
\font\ninerm=cmr9 
\font\ninei=cmmi9
\font\nineit=cmti9
\font\ninesl=cmsl9
\font\ninebf=cmbx9
\font\ninesy=cmsy9
\def\rmnine{\fam0\ninerm}
\def\itnine{\fam\itfam\nineit}
\def\slnine{\fam\slfam\ninesl}
\def\bfnine{\fam\bffam\ninebf}
\def\ninepoint{\let\rm=\rmnine
\textfont0=\ninerm \scriptfont0=\sevenrm \scriptscriptfont0=\fiverm
\textfont1=\ninei\scriptfont1=\seveni \scriptscriptfont1=\fivei
\textfont2=\ninesy
\textfont\itfam=\nineit \let\it=\itnine
\textfont\slfam=\ninesl \let\sl=\slnine
\textfont\bffam=\ninebf \scriptfont\bffam=\sevenbf
\scriptscriptfont\bffam=\fivebf
\let\bf=\bfnine
\let\sc=\sevenrm
\normalbaselineskip=11pt\normalbaselines\rm}
\font\tenib=cmmib10
\font\tensc=cmcsc10
\def\rmten{\fam0\tenrm}
\def\itten{\fam\itfam\tenit}
\def\slten{\fam\slfam\tensl}
\def\bften{\fam\bffam\tenbf}
\def\tenpoint{\let\rm=\rmten
\textfont0=\tenrm\scriptfont0=\sevenrm\scriptscriptfont0=\fiverm
\textfont1=\teni\scriptfont1=\seveni\scriptscriptfont1=\fivei
\textfont2=\tensy
\textfont\itfam=\tenit \let\it=\itten
\textfont\slfam=\tensl \let\sl=\slten
\textfont\bffam=\tenbf
\scriptfont\bffam=\sevenbf
\scriptscriptfont\bffam=\fivebf
\let\bf=\bften
\let\sc=\tensc
\normalbaselineskip=12pt\normalbaselines\rm}
\font\twelmib=cmmib10 scaled \magstep1
\font\twelbf=cmbx10   scaled \magstep1
\font\twelsy=cmsy10   scaled \magstep1
\def\rmbftwel{\fam0\twelbf}
\def\itbftwel{\fam\itfam\twelmib}
\def\bftwel{\fam\bffam\twelbf}
\def\bftwelpoint{\let\rm=\rmbftwel
\textfont0=\twelbf  \scriptfont0=\tenbf \scriptscriptfont0=\ninebf
\textfont1=\twelmib \scriptfont1=\tenib \scriptscriptfont1=\tenib
\textfont2=\twelsy
\textfont\itfam=\twelmib  \let\it=\itbftwel
\textfont\slfam=\twelmib  \let\sl=\itbftwel 
\textfont\bffam=\twelbf   \scriptfont\bffam=\tenbf
\scriptscriptfont\bffam=\ninebf
\let\bf=\bftwel
\normalbaselineskip=14pt\normalbaselines\rm}
\font\fortmib=cmmib10 scaled \magstep2
\font\fortbf=cmbx10   scaled \magstep2
\font\fortsy=cmsy10   scaled \magstep2
\def\rmbffort{\fam0\fortbf}
\def\itbffort{\fam\itfam\fortmib}
\def\bffort{\fam\bffam\fortbf}
\def\bffortpoint{\let\rm=\rmbffort
\textfont0=\fortbf  \scriptfont0=\twelbf \scriptscriptfont0=\tenbf
\textfont1=\fortmib \scriptfont1=\twelmib \scriptscriptfont1=\tenib
\textfont2=\fortsy
\textfont\itfam=\fortmib  \let\it=\itbffort
\textfont\slfam=\fortmib  \let\sl=\itbffort 
\textfont\bffam=\fortbf   \scriptfont\bffam=\twelbf
\scriptscriptfont\bffam=\tenbf
\let\bf=\bffort
\normalbaselineskip=16pt\normalbaselines\rm}
%
\def\ul#1{$\setbox0=\hbox{#1}\dp0=0pt\mathsurround=0pt
\underline{\box0}$}
%
\pageno=1
\parindent=4mm
\hsize=120mm \hoffset=20mm
\vsize=190mm \voffset=20mm
\outer\def\bye{\bigskip\vfill\supereject\end}
%
\def\raggedcenter{\leftskip=4em plus 12em \rightskip=\leftskip
  \parindent=0pt \parfillskip=0pt \spaceskip=.3333em \xspaceskip=.5em
  \pretolerance=9999 \tolerance=9999
  \hyphenpenalty=9999 \exhyphenpenalty=9999 }
\def\\{\break}
\newtoks\LECTURE \LECTURE={LECTURE}
\newcount\CNlecture \CNlecture=1
\long\def\title#1{\bgroup\vglue10mm\raggedcenter
	             {\tenpoint\bf\the\LECTURE\ \ArabicCN{lecture}}
	             \vskip10mm
	             {\bffortpoint #1}}
\long\def\author#1{\vskip5mm\tenpoint\rm #1}
\def\inst#1{$(^{#1})$}
\long\def\institute#1{\vskip5mm\tenpoint\sl#1}
\def\maketitle{\vskip5mm
               \vtop{\baselineskip=4pt
                     \vrule height.5pt width3cm\par
                     \vrule height.5pt width2cm}
               \vskip20mm\egroup\tenpoint
	          \let\lasttitle=Y\everypar={\let\lasttitle=N}}
%
\newbox\boxtitle
\newskip\beforesect \newskip\aftersect 
\newskip\beforesubsect \newskip\aftersubsect 
\newskip\beforesubsubsect \newskip\aftersubsubsect
%
\beforesect=7mm plus1mm minus1mm   \aftersect=5mm plus.5mm minus.5mm
\beforesubsect=5mm plus.5mm minus.5mm \aftersubsect=3mm plus.2mm minus.1mm 
\beforesubsubsect=3mm plus.2mm minus.1mm \aftersubsubsect=2mm plus.2mm minus.1mm 
\newcount\CNTa \CNTa=0
\newcount\CNTb \CNTb=0
\newcount\CNTc \CNTc=0
\newcount\CNTd \CNTd=0
\def\resetCN #1{\global\csname CN#1\endcsname =0}
\def\stepCN #1{\global
\expandafter\advance \csname CN#1\endcsname by 1}
\def\ArabicCN #1{\expandafter\number\csname CN#1\endcsname}
\def\RomanCN #1{\uppercase\expandafter{\romannumeral\csname CN#1\endcsname}}
\newcount\sectionpenalty  \sectionpenalty=0
\newcount\subsectionpenalty  \subsectionpenalty=0
\newcount\subsubsectionpenalty  \subsubsectionpenalty=0
\newdimen\indsect
\newdimen\dimensect
\indsect=1cm
\dimensect=\hsize\advance\dimensect by -\indsect
\def\section#1#2 {\par\resetCN{Tb}\resetCN{Tc}\resetCN{Td}%
              \if N\lasttitle\else\vskip-\beforesect\fi
              \bgroup
              \bf
              \pretolerance=20000
              \setbox0=\vbox{\vskip\beforesect
                       \noindent\ArabicCN{Ta}.\kern1em#1#2
                       \vskip\aftersect}
              \dimen0=\ht0\advance\dimen0 by\dp0 
              \advance\dimen0 by 2\baselineskip
              \advance\dimen0 by\pagetotal
              \ifdim\dimen0>\pagegoal
                 \ifdim\pagetotal>\pagegoal
                 \else\eject\fi\fi
              \vskip\beforesect
              \penalty\sectionpenalty \global\sectionpenalty=-200
              \global\subsectionpenalty=10007
              \ifx#1*\noindent #2\else\stepCN{Ta}
              \setbox0=\hbox{\noindent\ArabicCN{Ta}.}
              \indsect=\wd0\advance\indsect by 1em
              \parshape=2 0pt\hsize \indsect\dimensect
              \noindent\hbox to \indsect{\ArabicCN{Ta}.\hfil}#1\fi
              \vskip\aftersect
              \egroup
              \let\lasttitle=Y
              \nobreak\parindent=0pt
              \everypar={\parindent=4mm
                         \penalty0\let\lasttitle=N}\ignorespaces}
\def\subsection#1#2 {\par\resetCN{Tc}\resetCN{Td}%
              \if N\lasttitle\else\vskip-\beforesubsect\fi
              \bgroup\tenpoint\bf
                 \setbox0=\vbox{\vskip\beforesubsect
                 \noindent\ArabicCN{Ta}.\ArabicCN{Tb}.\kern1em#1#2
                 \vskip\aftersubsect}
              \dimen0=\ht0\advance\dimen0 by\dp0\advance\dimen0 by
                 2\baselineskip
              \advance\dimen0 by\pagetotal
              \ifdim\dimen0>\pagegoal
                 \ifdim\pagetotal>\pagegoal
                 \else \if N\lasttitle\eject\fi \fi\fi
              \vskip\beforesubsect
              \if N\lasttitle \penalty\subsectionpenalty \fi
              \global\subsectionpenalty=-100
              \global\subsubsectionpenalty=10007
              \ifx#1*\noindent#2\else\stepCN{Tb}
              \setbox0=\hbox{\noindent\ArabicCN{Ta}.\ArabicCN{Tb}.}
              \indsect=\wd0\advance\indsect by 1em
              \parshape=2 0pt\hsize \indsect\dimensect
              \noindent\hbox to \indsect{\ArabicCN{Ta}.\ArabicCN{Tb}.\hfil}#1\fi
              \vskip\aftersubsect
              \egroup\let\lasttitle=Y
              \nobreak\parindent=0pt
              \everypar={\parindent=4mm
                         \penalty0\let\lasttitle=N}\ignorespaces}
\def\subsubsection#1#2 {\par\resetCN{Td}%
              \if N\lasttitle\else\vskip-\beforesubsubsect\fi
              \bgroup\tenpoint\sl
                 \setbox0=\vbox{\vskip\beforesubsubsect\noindent
              {\ArabicCN{Ta}.\ArabicCN{Tb}.\ArabicCN{Tc}.\kern1em}#1#2
              \vskip\aftersubsubsect}
              \dimen0=\ht0\advance\dimen0 by\dp0\advance\dimen0 by
                 2\baselineskip
              \advance\dimen0 by\pagetotal
              \ifdim\dimen0>\pagegoal
                 \ifdim\pagetotal>\pagegoal
                 \else \if N\lasttitle\eject\fi \fi\fi
              \vskip\beforesubsubsect
              \if N\lasttitle \penalty\subsubsectionpenalty \fi
              \global\subsubsectionpenalty=-50
              \ifx#1*\noindent#2\else\stepCN{Tc}
              \setbox0=\hbox{\noindent
              	\ArabicCN{Ta}.\ArabicCN{Tb}.\ArabicCN{Tc}.}
              \indsect=\wd0\advance\indsect by 1em
              \parshape=2 0pt\hsize \indsect\dimensect
              \noindent\hbox to 
              \indsect{\ArabicCN{Ta}.\ArabicCN{Tb}.\ArabicCN{Tc}.\hfil}#1\/\fi
              \vskip\aftersubsubsect
              \egroup\let\lasttitle=Y
              \nobreak\parindent=0pt
              \everypar={\parindent=4mm
                         \penalty0\let\lasttitle=N}\ignorespaces}
\def\paragraph#1#2 {\par\if N\lasttitle\else\vskip-\aftersubsubsect\fi
    		    \bgroup\tenpoint\rm
         	    \setbox0=\vbox{\vskip\aftersubsubsect\noindent
         	    {\ArabicCN{Ta}.\ArabicCN{Tb}.\ArabicCN{Tc}.\ArabicCN{Td}}#1#2}
              \dimen0=\ht0\advance\dimen0 by\dp0\advance\dimen0 by
                 2\baselineskip
              \advance\dimen0 by\pagetotal
              \ifdim\dimen0>\pagegoal
              \ifdim\pagetotal>\pagegoal
              \else \if N\lasttitle\eject\fi \fi\fi
              \vskip\aftersubsubsect
              \if N\lasttitle \penalty-50 \fi
              \ifx#1*\noindent\ul{#2:}\ \else\stepCN{Td}
              \setbox0=\hbox{\noindent
              	\ArabicCN{Ta}.\ArabicCN{Tb}.\ArabicCN{Tc}.\ArabicCN{Td}.}
              \indsect=\wd0\advance\indsect by 1em
              \parshape=2 0pt\hsize \indsect\dimensect
              \noindent\hbox to 
              \indsect{\ArabicCN{Ta}.\ArabicCN{Tb}.\ArabicCN{Tc}.\ArabicCN{Td}.
              \hfil}\ul{#1:}\ \fi
              \egroup\let\lasttitle=N}
%
\newtoks\ACK \ACK={Acknowledgements}
\def\ack#1{\par\vskip\beforesect\goodbreak
\noindent{\tenpoint\bf\the\ACK }\par\vskip\aftersect\penalty500\noindent#1}
%
\newtoks\APPND \APPND={Appendix}
\def\appendix#1{\par\vskip\beforesect\goodbreak
\noindent{\tenpoint\bf\the\APPND \kern.5em#1}\par
\vskip\aftersect\penalty500\let\lasttitle=Y}
\def\titleapp #1{\if N\lasttitle\goodbreak
\else\vskip-\beforesubsect\penalty500\fi
\vskip\beforesubsect
{\tenpoint\bf\noindent\ignorespaces #1}
\vskip\aftersubsect\let\lasttitle=Y\noindent}
%
\newtoks\REFNAME \REFNAME={References}
\def\references{\begREF}
\def\begREF{\bgroup
              \setbox0=\vbox{\vskip\beforesect\noindent{\bf\the\REFNAME}
                       \vskip\aftersect}
              \dimen0=\ht0\advance\dimen0 by\dp0 
              \advance\dimen0 by 2\baselineskip
              \advance\dimen0 by\pagetotal
              \ifdim\dimen0>\pagegoal
                 \ifdim\pagetotal>\pagegoal
                 \else\eject\fi\fi
              \vskip\beforesect\noindent{\bf\the\REFNAME}
              \vskip\aftersect
               \frenchspacing \parindent=0pt \leftskip=1truecm
               \everypar{\hangindent=\parindent}}
\def\ref#1{ $[{\rm #1}]$}%
\gdef\refis#1{\item{$[$#1$]$\ }}   
\def\endreferences{\par\egroup}
%
\def\review#1, #2, 1#3#4#5, #6 {{\sl#1\/} {\bf#2} (1#3#4#5) #6}
\def\book#1, #2, #3, 1#4#5#6, #7 {#1 (#2, #3, 1#4#5#6) p. #7}
%
\newcount\foCN \foCN=0
\def\fonote{\global\advance\foCN by 1
$(^{\rm\number\foCN})$\vfootnote{$(^{\rm\number\foCN})$}}
%
\catcode`@=11
\def\vfootnote#1{\insert\footins\bgroup
  \ninepoint
  \interlinepenalty\interfootnotelinepenalty
  \splittopskip\ht\strutbox 
  \splitmaxdepth\dp\strutbox \floatingpenalty\@MM
  \leftskip\z@skip \rightskip\z@skip \spaceskip\z@skip \xspaceskip\z@skip
  \baselineskip=10pt\lineskip=10pt
  \noindent\kern10mm\llap{#1\enspace}\footstrut
  \futurelet\next\fo@t}
\def\@foot{\egroup}
\catcode`@=12
%
\newtoks\shorttitle
\newtoks\authors
\shorttitle={SHORT TITLE}%
\authors={The Authors}%
\def\firsthd{\hfill}
\def\hdleft{\tenpoint\folio\hfill{\sl \the\authors\/}\hfill}
\def\hdrigt{\tenpoint\hfill{\ninepoint\the\shorttitle}\hfill\folio}
\newif\ifbegpage
\headline={\ifbegpage\firsthd
\global\begpagefalse\else
\ifodd\pageno\hdrigt\else\hdleft\fi
\advance\pageno by 1\fi}
\footline={\hfil}
\def\makeheadline{\vbox to0pt{\vskip-10mm
\line{\vbox to8.5pt{}\the\headline}\vss}\nointerlineskip}
%
\begpagetrue
%
\newtoks\TABLE \TABLE={Table}
%
\newdimen\tableheight
\newskip\superskipamount \superskipamount=28pt plus 4pt minus 4pt
\def\superskip{\vskip\superskipamount}
\def\superbreak{\par\ifdim\lastskip<\superskipamount
  \removelastskip\penalty-200\superskip\fi}
\catcode`\@=11
\def\endinsert{\egroup 
  \if@mid \dimen@\ht\z@ \advance\dimen@\dp\z@
    \advance\dimen@24\p@ \advance\dimen@\pagetotal
    \ifdim\dimen@>\pagegoal\@midfalse\p@gefalse\fi\fi
  \if@mid \superskip\box\z@\superbreak
  \else\insert\topins{\penalty100 
    \splittopskip\z@skip
    \splitmaxdepth\maxdimen \floatingpenalty\z@
    \ifp@ge \dimen@\dp\z@
    \vbox to\vsize{\unvbox\z@\kern-\dimen@}
    \else \box\z@\nobreak\superskip\fi}\fi\endgroup}
\catcode`\@=12
\newcount\CNfig \CNfig=0
\let\captext=N
\def\begfig #1cm{\midinsert\tableheight=#1cm\advance\tableheight by 5mm
	\vglue\tableheight}
\def\topfig #1cm{\topinsert\tableheight=#1cm\advance\tableheight by 5mm
	\vglue\tableheight}
\long\def\caption#1{\let\captext=Y\stepCN{fig}\ninepoint\rm
	\noindent Fig.\ \number\CNfig.\kern.3em ---\kern.3em\ignorespaces
     \parindent=0pt#1\par}
\def\endfig{\if N\captext\stepCN{fig}
	\ninepoint\rm\noindent Figure \number\CNfig\else\let\figtext=N\fi
	\endinsert}
%
\newcount\CNtab \CNtab=0
\let\tabtext=N
\def\begtab #1cm{\midinsert\tableheight=#1cm}
\def\toptab #1cm{\topinsert\tableheight=#1cm}
\long\def\tabcap#1{\let\tabtext=Y\stepCN{tab}
	\ninepoint\rm\noindent Table\ \RomanCN{tab}.\kern.3em ---\kern.3em\ignorespaces
	\parindent=0pt#1\par}
\def\endtab{\if N\tabtext\stepCN{tab}
	\ninepoint\rm\noindent Table \RomanCN{tab} \else\let\tabtext=N\fi
	\ifdim\tableheight=0cm\vskip-\belowdisplayskip
     \else\advance\tableheight by 5mm\fi
     \vglue\tableheight\endinsert}
%

%
\def\(#1){(\call{#1})}

\catcode`@=11
\newcount\r@fcount \r@fcount=0
\newcount\r@fcurr
\immediate\newwrite\reffile
\newif\ifr@ffile\r@ffilefalse
\def\w@rnwrite#1{\ifr@ffile\immediate\write\reffile{#1}\fi\message{#1}}
 
\def\writer@f#1>>{}
\def\referencefile{
  \r@ffiletrue\immediate\openout\reffile=\jobname.ref%
  \def\writer@f##1>>{\ifr@ffile\immediate\write\reffile%
    {\noexpand\refis{##1} = \csname r@fnum##1\endcsname = %
     \expandafter\expandafter\expandafter\strip@t\expandafter%
     \meaning\csname r@ftext\csname r@fnum##1\endcsname\endcsname}\fi}%
  \def\strip@t##1>>{}}

\def\citeall#1{\xdef#1##1{#1{\noexpand\cite{##1}}}}
\def\cite#1{\each@rg\citer@nge{#1}}
 
\def\each@rg#1#2{{\let\thecsname=#1\expandafter\first@rg#2,\end,}}
\def\first@rg#1,{\thecsname{#1}\apply@rg}
\def\apply@rg#1,{\ifx\end#1\let\next=\relax
\else,\thecsname{#1}\let\next=\apply@rg\fi\next}
 
\def\citer@nge#1{\citedor@nge#1-\end-}
\def\citer@ngeat#1\end-{#1}
\def\citedor@nge#1-#2-{\ifx\end#2\r@featspace#1 
  \else\citel@@p{#1}{#2}\citer@ngeat\fi}
\def\citel@@p#1#2{\ifnum#1>#2{\errmessage{Reference range #1-#2\space is bad.}%
    \errhelp{If you cite a series of references by the notation M-N, then M and
    N must be integers, and N must be greater than or equal to M.}}\else%
 {\count0=#1\count1=#2\advance\count1 by1\relax\expandafter\r@fcite\the\count0,%
  \loop\advance\count0 by1\relax
    \ifnum\count0<\count1,\expandafter\r@fcite\the\count0,%
  \repeat}\fi}
 
\def\r@featspace#1#2 {\r@fcite#1#2,}
\def\r@fcite#1,{\ifuncit@d{#1}
    \newr@f{#1}%
    \expandafter\gdef\csname r@ftext\number\r@fcount\endcsname%
                     {\message{Reference #1 to be supplied.}%
                      \writer@f#1>>#1 to be supplied.\par}%
 \fi%
 \csname r@fnum#1\endcsname}
\def\ifuncit@d#1{\expandafter\ifx\csname r@fnum#1\endcsname\relax}%
\def\newr@f#1{\global\advance\r@fcount by1%
    \expandafter\xdef\csname r@fnum#1\endcsname{\number\r@fcount}}
 
\let\r@fis=\refis
\def\refis#1#2#3\par{\ifuncit@d{#1}
   \newr@f{#1}%
   \w@rnwrite{Reference #1=\number\r@fcount\space is not cited up to now.}\fi%
  \expandafter\gdef\csname r@ftext\csname r@fnum#1\endcsname\endcsname%
  {\writer@f#1>>#2#3\par}}
 
\def\ignoreuncited{
   \def\refis##1##2##3\par{\ifuncit@d{##1}%
     \else\expandafter\gdef\csname r@ftext\csname r@fnum##1\endcsname\endcsname%
     {\writer@f##1>>##2##3\par}\fi}}
 
\def\r@ferr{\endreferences\errmessage{I was expecting to see
\noexpand\endreferences before now;  I have inserted it here.}}
\let\r@ferences=\references
\def\references{\r@ferences\def\endmode{\r@ferr\par\endgroup}}
 
\let\endr@ferences=\endreferences
\def\endreferences{\r@fcurr=0
  {\loop\ifnum\r@fcurr<\r@fcount
    \advance\r@fcurr by 1\relax\expandafter\r@fis\expandafter{\number\r@fcurr}%
    \csname r@ftext\number\r@fcurr\endcsname%
  \repeat}\gdef\r@ferr{}\endr@ferences}
 
 
\let\r@fend=\endpaper\gdef\endpaper{\ifr@ffile
\immediate\write16{Cross References written on []\jobname.REF.}\fi\r@fend}
 
\catcode`@=12

\citeall\ref%

\catcode`@=11
\newcount\tagnumber\tagnumber=0
 
\immediate\newwrite\eqnfile
\newif\if@qnfile\@qnfilefalse
\def\write@qn#1{}
\def\writenew@qn#1{}
\def\w@rnwrite#1{\write@qn{#1}\message{#1}}
\def\@rrwrite#1{\write@qn{#1}\errmessage{#1}}
 
\def\taghead#1{\gdef\t@ghead{#1}\global\tagnumber=0}
\def\t@ghead{}
 
\expandafter\def\csname @qnnum-3\endcsname
  {{\t@ghead\advance\tagnumber by -3\relax\number\tagnumber}}
\expandafter\def\csname @qnnum-2\endcsname
  {{\t@ghead\advance\tagnumber by -2\relax\number\tagnumber}}
\expandafter\def\csname @qnnum-1\endcsname
  {{\t@ghead\advance\tagnumber by -1\relax\number\tagnumber}}
\expandafter\def\csname @qnnum0\endcsname
  {\t@ghead\number\tagnumber}
\expandafter\def\csname @qnnum+1\endcsname
  {{\t@ghead\advance\tagnumber by 1\relax\number\tagnumber}}
\expandafter\def\csname @qnnum+2\endcsname
  {{\t@ghead\advance\tagnumber by 2\relax\number\tagnumber}}
\expandafter\def\csname @qnnum+3\endcsname
  {{\t@ghead\advance\tagnumber by 3\relax\number\tagnumber}}
 
\def\equationfile{%
  \@qnfiletrue\immediate\openout\eqnfile=\jobname.eqn%
  \def\write@qn##1{\if@qnfile\immediate\write\eqnfile{##1}\fi}
  \def\writenew@qn##1{\if@qnfile\immediate\write\eqnfile
    {\noexpand\tag{##1} = (\t@ghead\number\tagnumber)}\fi}
}
 
\def\callall#1{\xdef#1##1{#1{\noexpand\call{##1}}}}
\def\call#1{\each@rg\callr@nge{#1}}
 
\def\each@rg#1#2{{\let\thecsname=#1\expandafter\first@rg#2,\end,}}
\def\first@rg#1,{\thecsname{#1}\apply@rg}
\def\apply@rg#1,{\ifx\end#1\let\next=\relax%
\else,\thecsname{#1}\let\next=\apply@rg\fi\next}
 
\def\callr@nge#1{\calldor@nge#1-\end-}
\def\callr@ngeat#1\end-{#1}
\def\calldor@nge#1-#2-{\ifx\end#2\@qneatspace#1 %
  \else\calll@@p{#1}{#2}\callr@ngeat\fi}
\def\calll@@p#1#2{\ifnum#1>#2{\@rrwrite{Equation range #1-#2\space is bad.}
\errhelp{If you call a series of equations by the notation M-N, then M and
N must be integers, and N must be greater than or equal to M.}}\else%
 {\count0=#1\count1=#2\advance\count1 by1\relax\expandafter\@qncall\the\count0,%
  \loop\advance\count0 by1\relax%
    \ifnum\count0<\count1,\expandafter\@qncall\the\count0,%
  \repeat}\fi}
 
\def\@qneatspace#1#2 {\@qncall#1#2,}
\def\@qncall#1,{\ifunc@lled{#1}{\def\next{#1}\ifx\next\empty\else
  \w@rnwrite{Equation number \noexpand\(>>#1<<) has not been defined yet.}
  >>#1<<\fi}\else\csname @qnnum#1\endcsname\fi}
 
\let\eqnono=\eqno
\def\eqno(#1){\tag#1}
\def\tag#1$${\eqnono(\displayt@g#1 )$$}
 
\def\aligntag#1\endaligntag
  $${\gdef\tag##1\\{&(##1 )\cr}\eqalignno{#1\\}$$
  \gdef\tag##1$${\eqnono(\displayt@g##1 )$$}}

\def\eqalignno#1{\displ@y \tabskip\centering
  \halign to\displaywidth{\hfil$\displaystyle{##}$\tabskip\z@skip
    &$\displaystyle{{}##}$\hfil\tabskip\centering
    &\llap{$\displayt@gpar##$}\tabskip\z@skip\crcr
    #1\crcr}}
 
\def\displayt@gpar(#1){(\displayt@g#1 )}
 
\def\displayt@g#1 {\rm\ifunc@lled{#1}\global\advance\tagnumber by1
        {\def\next{#1}\ifx\next\empty\else\expandafter
        \xdef\csname @qnnum#1\endcsname{\t@ghead\number\tagnumber}\fi}%
  \writenew@qn{#1}\t@ghead\number\tagnumber\else
        {\edef\next{\t@ghead\number\tagnumber}%
        \expandafter\ifx\csname @qnnum#1\endcsname\next\else
        \w@rnwrite{Equation \noexpand\tag{#1} is a duplicate number.}\fi}%
  \csname @qnnum#1\endcsname\fi}
 
\def\ifunc@lled#1{\expandafter\ifx\csname @qnnum#1\endcsname\relax}
 
\let\@qnend=\end\gdef\end{\if@qnfile
\immediate\write16{Equation numbers written on []\jobname.EQN.}\fi\@qnend}
 
\catcode`@=12
 
 

\input epsf

\authors{James P. Sethna, Olga Perkovi\'{c}, and Karin A. Dahmen}
\shorttitle{Hysteresis, Avalanches, ...}

%
\CNlecture=10
%
\title{Hysteresis, Avalanches, and Barkhausen Noise}
\author{James P. Sethna\inst{1}, \\
Olga Perkovi\'{c}\inst{1}, and
Karin A. Dahmen\inst{2}}
\institute{\inst{1} Laboratory of Atomic and Solid State Physics\\
Cornell University, Ithaca, NY 14853-2501, USA

\inst{2} Department of Physics\\
Harvard University, Cambridge, MA 02138, USA
}

\maketitle

\section{INTRODUCTION}

We've been working on the crackling noise in hysteresis
loops\ref{prl1, prl2, prl3, prb,  perkovic, new}.
Hysteresis occurs when you push and pull on a system with an external
force, and the response lags behind the force.  The hysteresis loop is
the graph of force (say, an external magnetic field $H$) versus
the response (say, the magnetization $M$ of the material).
In many materials, the hysteresis loop is not actually microscopically smooth:
it is composed of small bursts, or avalanches (figure~1).  In many
first-order phase transitions, these bursts cause acoustic emission
(crackling noise); in magnets, they are called Barkhausen noise.

\midinsert
\centerline{
\epsfxsize=2.2truein
\epsffile{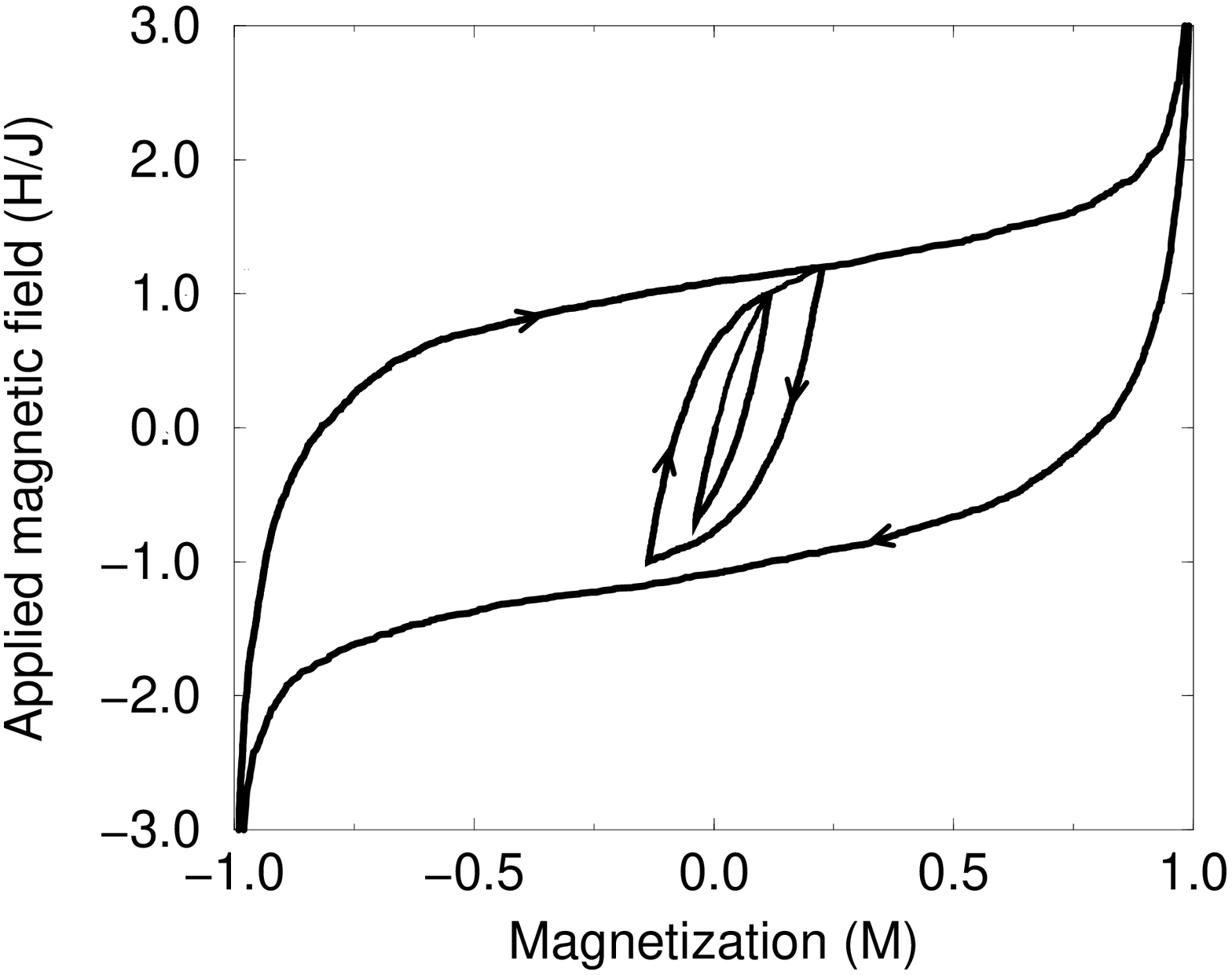}
\epsfxsize=2.2truein
\epsffile{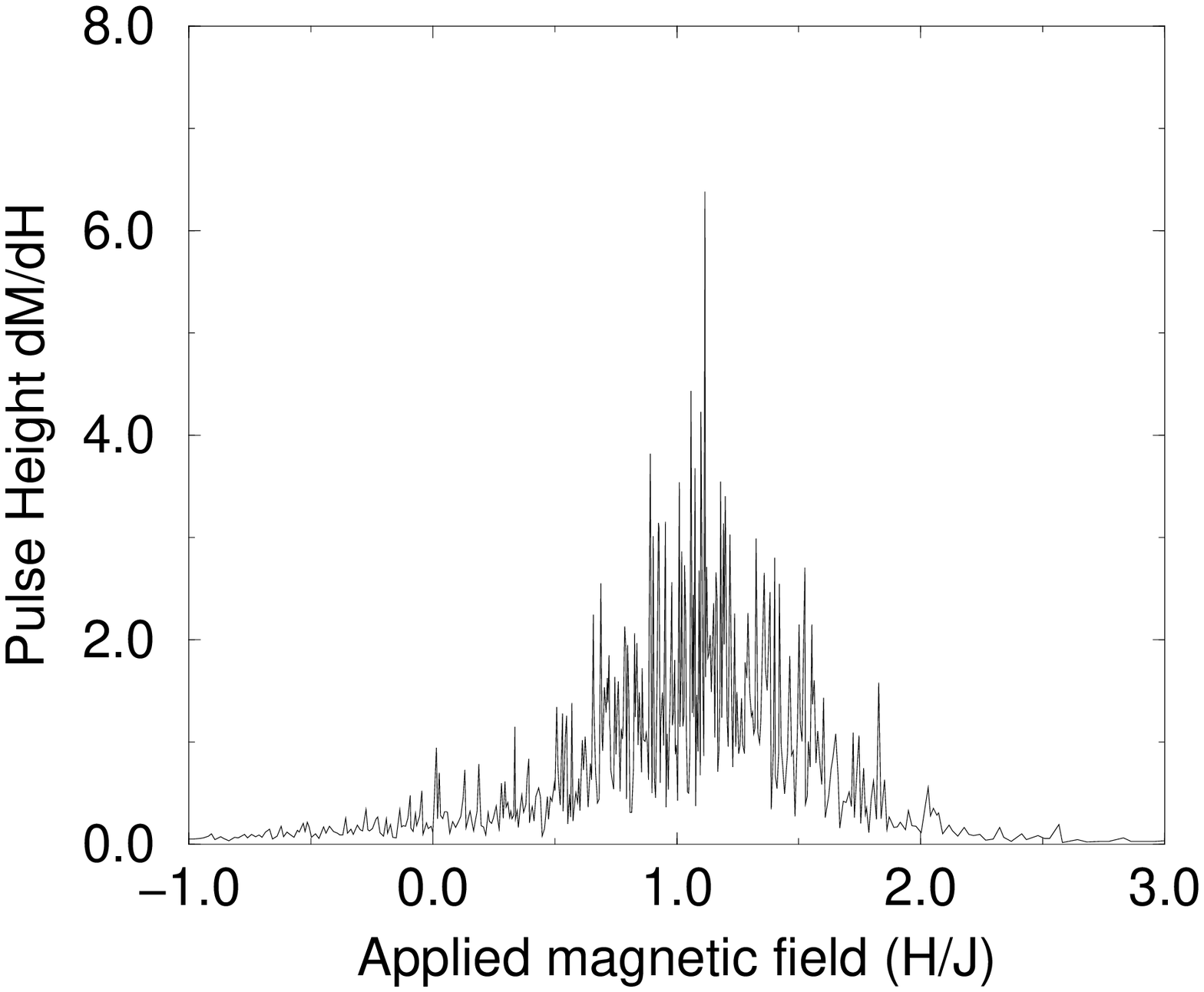}
}
\caption{Left:~A hysteresis loop in our model, showing the subloops.
If you look carefully, you should be able to see small irregularities in the
curve: these correspond to the avalanches causing the crackling noise.
Right:~The pulses in the upper branch of the outer hysteresis loop.
Notice that the pulses become larger near $H=1$ where the magnetization
changes fastest with external field.}
\endinsert

Naturally, these pulses are associated with some kind of inhomogeneity
or disorder in the material.  Magnetic tapes are composed of small grains of
iron oxide, and individual grains (when small enough) flip over as a
unit, leading to a pulse in the magnetization.  However, the pulses
observed have a wide range of sizes: they can range over three to six decades
in size in a typical experiment (figure~2).  Since the grains in the
material don't come in such a variety of sizes, one can conclude
that many grains must be flipping at once, coupled together in a kind of
avalanche.

\midinsert
\centerline{
\epsfxsize=2.2truein
\epsffile{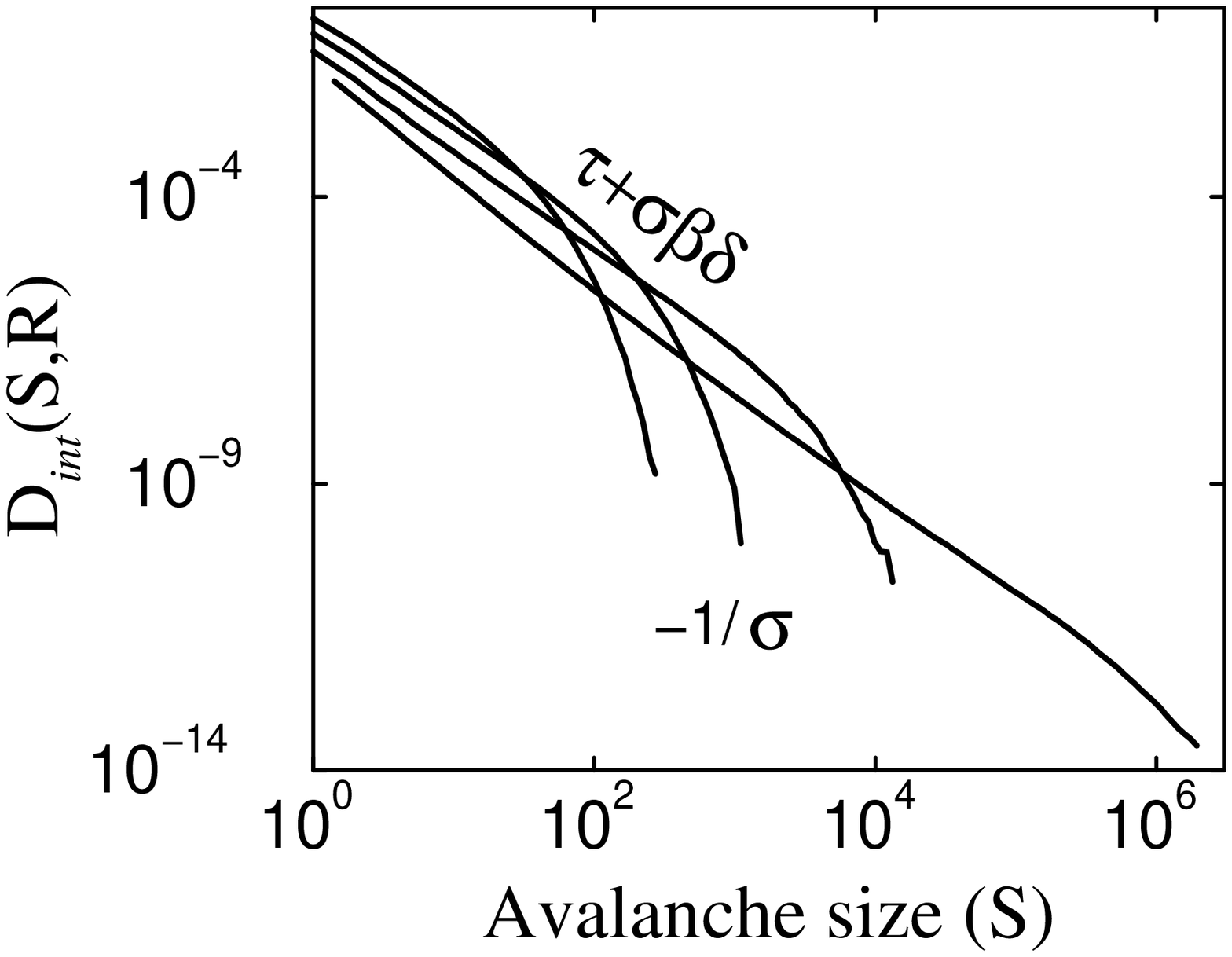}
\epsfxsize=2.2truein
\epsffile{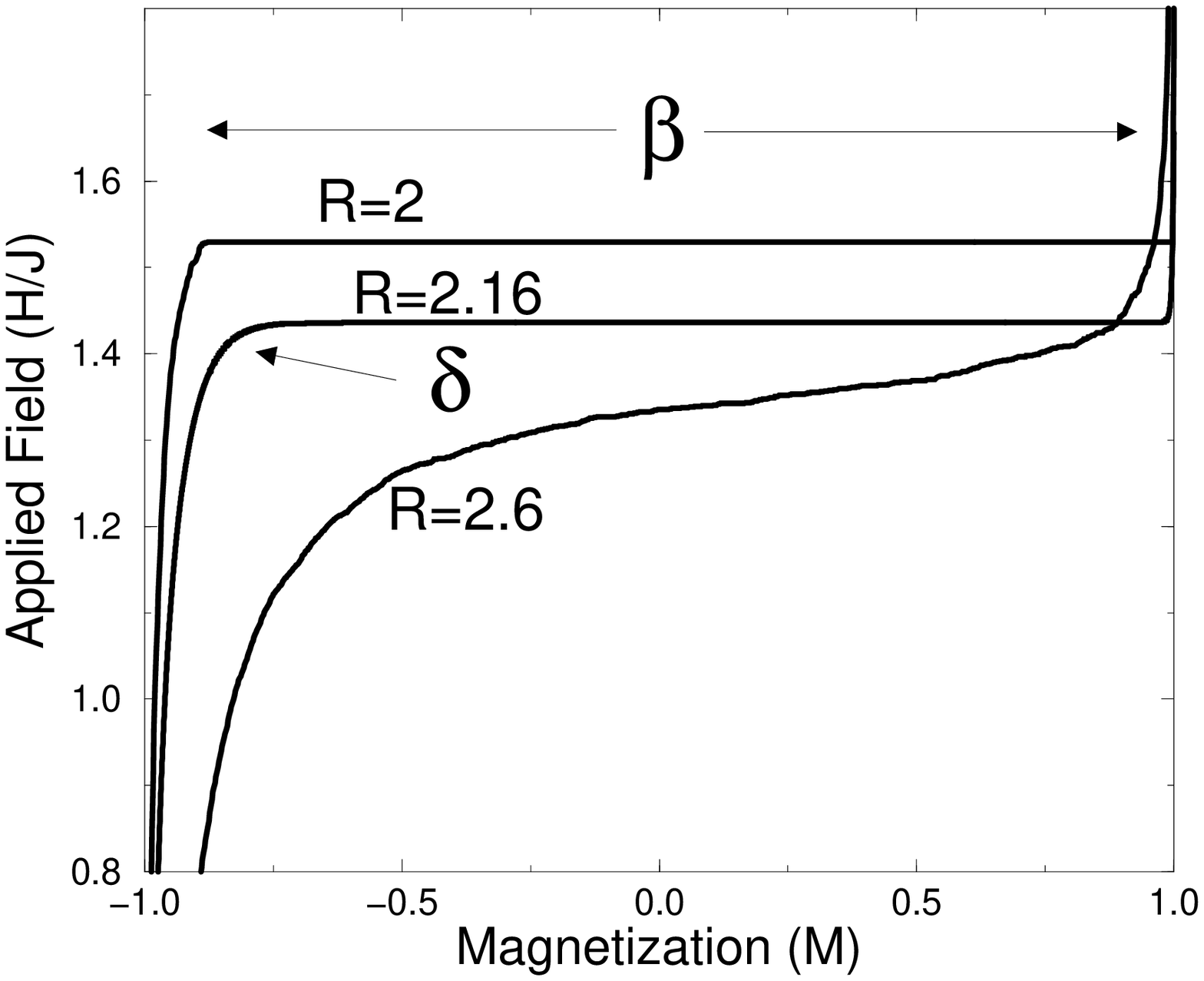}
}
\caption{Left:~The avalanche size distribution for our model in three
dimensions, for disorders $R = 4$, 3.2, 2.6, and 2.25.  The dashed line
shows the expected behavior at the critical disorder $R_c\sim 2.16$
(a power law, $D_{int}(s, R_c)\propto s^{-(\tau + \sigma\beta\delta)}$).
The curves are cut off after a few decades of scaling, at a size
$S_{cutoff} \propto (R-R_c)^{-1/\sigma}$.
Right:~The jump in the avalanche size distribution ends at the critical
disorder $R_c=2.16$.  The size of the jump scales as
$\Delta M \propto (R-R_c)^\beta$.  At $R_c$, the magnetization
has a power-law form $H(M) \propto (M-M_c)^\delta$.
}
\endinsert

Having events of all sizes is not trivial!  If the coupling between
grains is weak compared to the disorder, the grains will tend to flip
independently, leading to small avalanches.  If the coupling is strong,
a grain which flips will give a large kick to its neighbors, likely
flipping several of them, who will flip several more --- leading to
one large avalanche.  On the right in figure 2, we see that this is
precisely what happens in our model.  The large range of avalanches
is associated with a critical value of the disorder $R_c$ relative
to the coupling: when the avalanches can't decide whether to be huge
or small, they come in all sizes!

\section{THE MODEL}

To model these magnetic systems, we use a lattice of ``spins'' $S_i=\pm 1$,
pointing up or down: each spin represents a domain or particle in the
material.  They are attached to their nearest neighbor (n.n.) spins by bonds
of uniform strength $J$ which we set for convenience to one; they are
coupled to an external field $H(t)$ which we sweep from $-\infty$ to $\infty$.
Finally, we model the inhomogeneities in the material with a random bias for
up or down: at each site, we pick a random field $h_i$, distributed
with a normal probability distribution $P(h)=\exp(-h^2/2 R^2)/\sqrt{2 \pi}R$.
We call $R$ the disorder: large $R$ makes the distribution of $h_i$ broad,
makes the coupling between spins unimportant, and leads to smooth hysteresis
loops and small avalanches.  The energy of our system thus is
$$H=-\sum_{ij~n.n.} J S_i S_j - \sum_i (H(t)+h_i) s_i. \tag(Hamiltonian)$$
Each spin flips as soon as its local external field $J\sum_{n.n.}s_j+H(t)+h_i$
changes sign: it then can kick over its neighbors if the resulting $2J$
change in their local fields is big enough.  Thermal (and quantum)
fluctuations aren't important for us, because the particles are large
and (often by design) don't flip over spontaneously.  Our model
is called the random-field Ising model, and we simulate it at zero
temperature \ref{depinning, simulations}.

\section{THE CRITICAL EXPONENTS}

It's a remarkable fact that models like ours can accurately describe
real systems near their transitions.  The basic idea is a lot like
hydrodynamics.  All kinds of fluids look alike at long lengths and times,
apart from their viscosity, density, and surface tension: despite
rather different molecular structures and interactions, they all
are described by the same equations for long distances and long times
compared to the atomic scales.  Similarly, a variety of hysteretic
systems near the onset of a big jump (the ``infinite avalanche'') should
be quantitatively describable by our simple model.  This amazing property
is called {\it universality}, and the family of models with common
descriptions is called the {\it universality class}.

The most commonly measured universal quantities are the critical exponents.
Many properties near the critical point have power-law scaling.  This
can be understood as a kind of self-universality: the system on one
scale is quantitatively described by the same system at a different
scale.  Thus the properties of the system become scale invariant, and
(in the usual way) develop power laws.

There are several critical exponents for our system that we measure
and calculate.  The most common is the power law giving the number of
avalanches of a given size.  This power law depends on whether you
count all the avalanches in the hysteresis loop (the integrated
avalanche size distribution plotted on the left in figure~2, whose power law
is $\tau + \sigma \beta \delta$), or just measure them near the
incipient jump in $M(H)$ (near which point the power law is $\tau$).
The exponents $\beta$, $\delta$, and $\sigma$ are also described
in figure~2: they describe the shapes of the hysteresis loops and
the cutoff in the avalanche size distribution as the disorder $R$ is varied.
The exponent $\nu$ describes the dependence of the size of the large
avalanches on the distance from the critical point; the exponent $z$
describes the lifetime of the large avalanches.

The first two columns of the left of figure~3 compare the two
avalanche-size power-laws with experiments on a variety of 
materials\ref{experiments}.  While the scatter is large, the theory is well
within the range of exponents measured.   That doesn't mean we know the
experiments are described by our theory: there might well be other
universality classes with exponents not so far from ours...  The other
columns on the left of figure~3 represent combinations of exponents
derived from different kinds of
measurements: for example, the power spectrum of the noise
from the hysteresis loops.

\midinsert
\centerline{
\epsfxsize=2.2truein
\epsffile{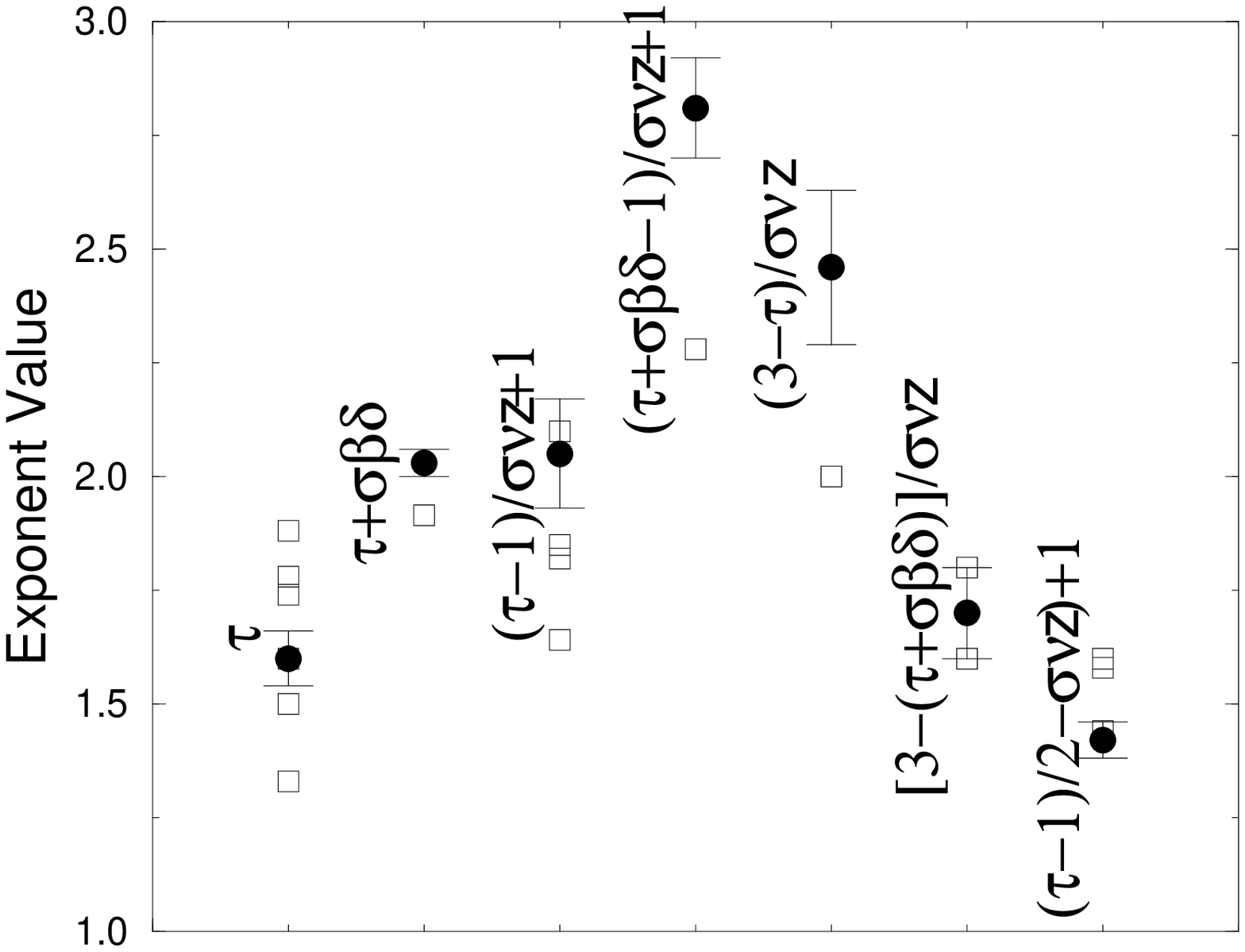}
\epsfxsize=2.2truein
\epsffile{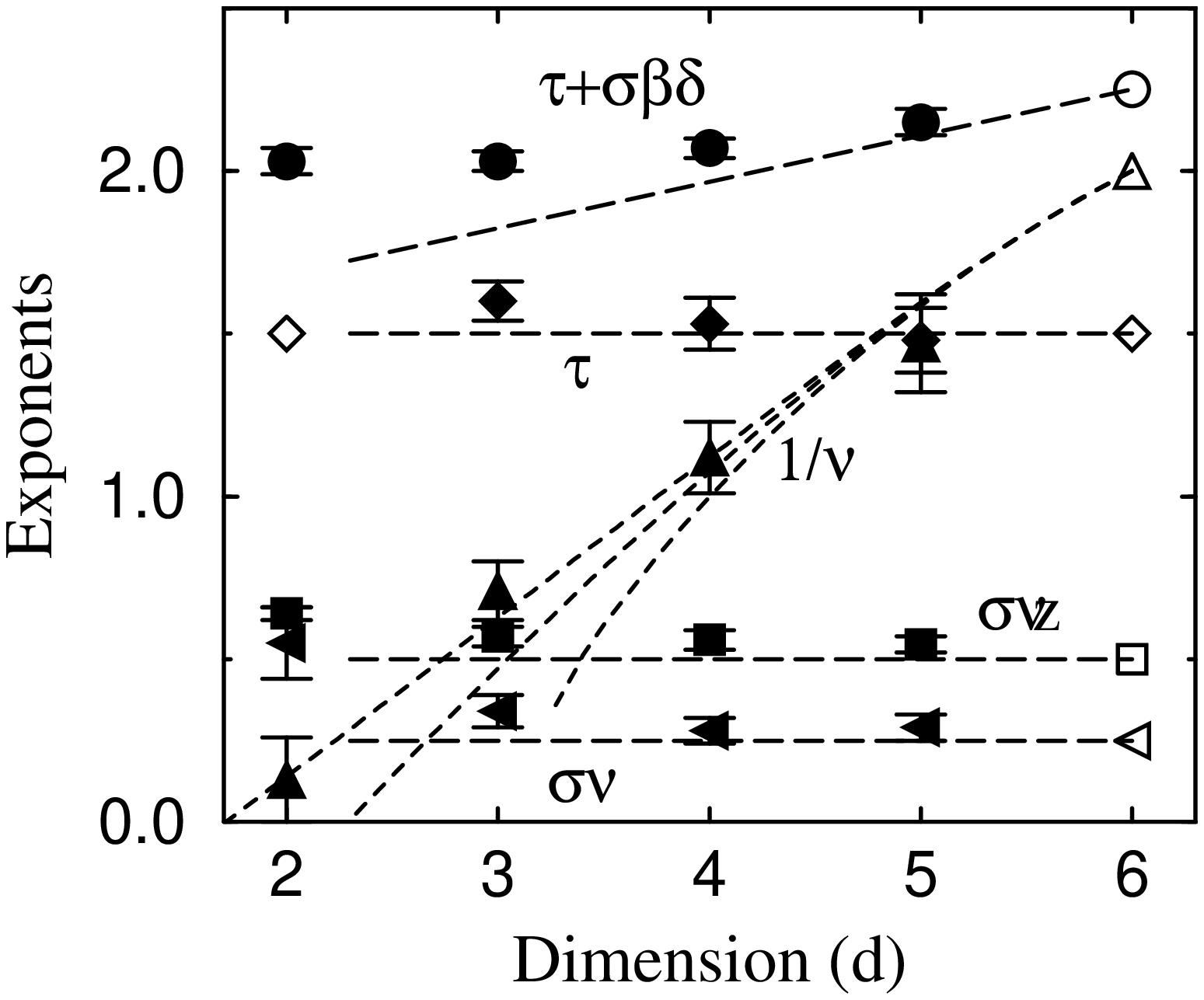}
}
\caption{Left:~Comparison of critical exponents with a variety of
experiments \ref{experiments}, from \ref{prl3}.
Right:~Exponents in various dimensions \ref{prl3}.  The dashed lines are
from the $\epsilon$-expansion \ref{prl2, prb, epsilon5}.
}
\endinsert

On the right-hand side of figure~3 we see the critical exponents in
different spatial dimensions.  Two dimensions might describe the
behavior of magnetic tapes.  Of course, dimensions greater than three
have only theoretical interest!  We'll see that we can learn something
from high dimensions anyway...

\section{THE EPSILON EXPANSION}

What justifies us in thinking that our exponents and scaling is universal?
How can we explain why we expect many systems to have exactly the same critical
exponents (and scaling functions, and amplitude ratios, ...)?  The
theoretical justification of this was given by Leo Kadanoff, Ken Wilson,
and Michael Fisher in the early 1970's, using what is called the
{\it renormalization group}\ref{epsilon}.  It's a bit technical, but theorists
get unhappy unless they can point to a difficult calculation underlying
their work.

The basic idea of the renormalization group is to think of the process
of rescaling a system as a mapping from the space of all systems to itself!
In statistical mechanics, one describes a system with a Hamiltonian:
coarse-graining from one length scale to another maps Hamiltonian space
into itself.  In studying the period-doubling route for the onset of chaos,
one describes the system with a mapping: rescaling the time by a factor
of two is done by composing the mapping with itself.  In our problem,
we write a path integral for the probability of all histories for the
system, and consider the effect of coarse-graining from one length-scale
to another as a mapping taking one path-probability functional onto
another\ref{epsilon}.

This leap of abstraction is more useful than you might imagine.
The subspace of all systems at their critical points must map onto
itself.  (If a system is on the verge of having an infinite avalanche,
looking at it on a longer length scale won't change that.)  Suppose one
of the critical systems is a fixed-point under the mapping: suppose
it has a ``basin of attraction'' of critical systems which flow towards
it under coarse-graining.  Then, on long length scales, all of these
systems should look like the fixed point: the basin of attraction becomes
the universality class.

\midinsert
\centerline{
\epsfxsize=4.4truein
\epsffile{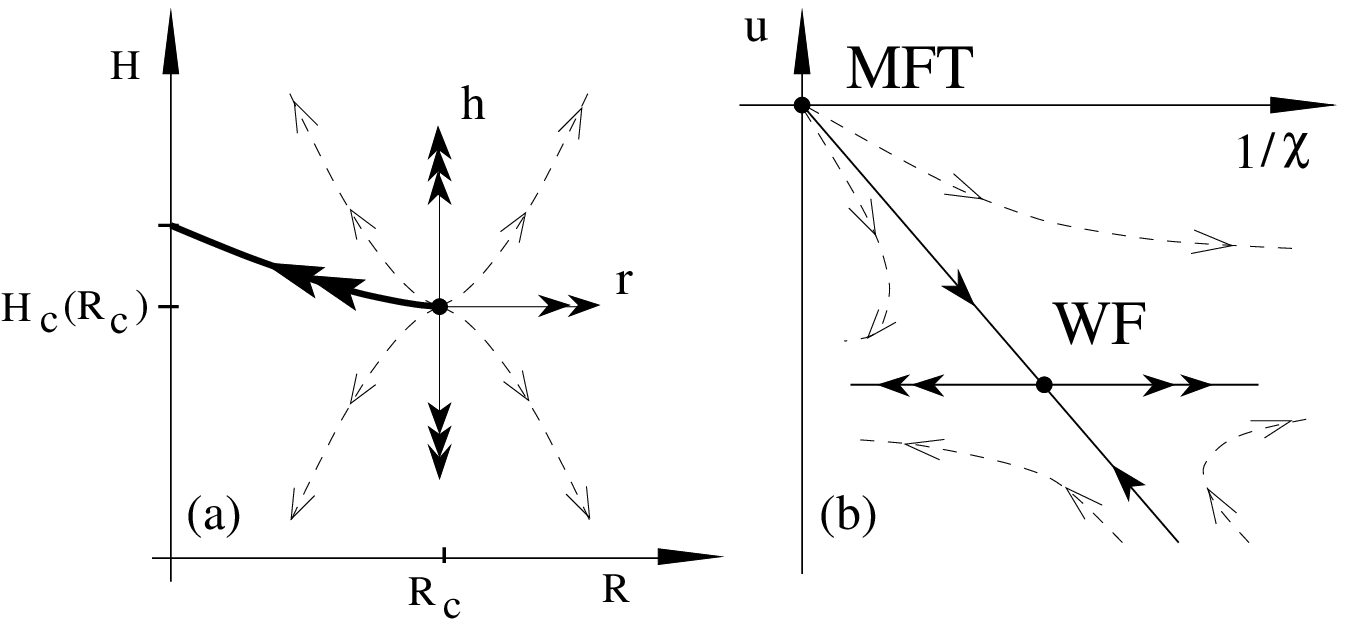}
}
\caption{Left:~Schematic phase diagram for our model, with arrows showing
flows under coarse-graining.  The dark line is $H_c(R)$, the external field
at which the infinite avalanche occurs when the system is swept upwards
from $H=-\infty$.  Under coarse-graining, the effective external field
$h = (H-H_c)/H$ grows fastest, and the effective disorder $r=(R-R_c)/R$
grows more slowly: all other directions are stable under coarse-graining.
Right:~Flow diagram near six dimensions, showing the mean-field fixed-point
and the new, Wilson-Fisher fixed point.  The mean-field fixed point is
unstable below $d=6$; at six dimensions, these two points merge.
}
\endinsert

Karin Dahmen\ref{prl2, prb} figured out how to implement this in a
tangible calculation. In dimension $d$, each spin on a hypercubic
lattice has $2d$ neighbors: in high dimensions, there are so many
neighbors that one may just as well assume that every spin sees an
average environment.  This {\it mean-field theory} can be solved; it has
a transition where the infinite avalanche first occurs, and that point
leads to a fixed-point under the rescaling. In high dimensions (for us,
dimensions $d>6$) this fixed point is stable: all systems have the same
critical properties as the mean-field fixed point. As the dimension
decreases below $d=6$, the mean-field fixed-point becomes unstable, but
Karin did a perturbation theory in the dimension $\epsilon = 6-d$ to
find the new fixed-point to first order in $\epsilon$ (figure~4).  The
calculation is completely analogous to Ken Wilson's original calculation
of the thermal, pure Ising model in $d = 4-\epsilon$.

The results of Karin's calculation are shown in the right-hand part of
figure~3.  It's nice to see that the critical exponents approach their
mean-field values as $d$ approaches six, and that the $\epsilon$-expansion
captures the first corrections rather well.  Indeed, the method works
amazingly well, considering $\epsilon \ge 3$ in realistic problems!
It so happens that the analogy to Ken Wilson's calculation is embarrassingly
good: our calculation agrees with his to all orders in $\epsilon$.
Other people have calculated terms up to order $\epsilon^5$ for
the exponent $\nu$ in the pure, thermal Ising model\ref{epsilon5}; we use their
coefficients (in two higher dimensions) to predict $1/\nu$ in figure 3 right.
There is one subtlety: the series for $\nu(\epsilon)$ doesn't converge 
without help (it's an asymptotic series), so we plot three different 
Borel resummations of the series for $1/\nu$ \ref{epsilon5}.

The fact that our series in $6-\epsilon$ maps to Wilson's calculation
in $4-\epsilon$ is embarrassing for another reason.  Our values
for $\nu$ in $d=3$ definitely do not agree with the pure, thermal
Ising model in $d=1$!  This caused great anguish when it was first
discovered in another context\ref{epsilon5}.  It's of course possible that we've
erred in trying to perturb in a discrete variable like the dimension.
The method seems to be working rather well, though, from figure 3.
I personally believe it has something to do with the fact that the
series doesn't converge for either problem: maybe we have one series
trying to describe two different functions\ref{prl3,prb}?

\section{WIDOM SCALING}

It's important to stress that critical exponents are not the only
predictions of the theory.  I'd like to briefly discuss Ben Widom's
discovery (later explained using the renormalization group): data
for systems near criticality can be collapsed onto one another.

Consider again the avalanche size distribution (left figure~5, the same
data as in left figure~2).  Notice how the curves never quite lie along the
dashed line, which I claimed was the power-law you would see if you
were exactly at the critical point.  Even when there are avalanches of
size $10^6$, the system still isn't exhibiting the critical exponents
predicted!  No wonder the experiments on the left of figure~3 don't
agree with our theory: if you fit a power law to three decades of data,
the exponent you get depends on $R-R_c$ and on which three decades you
measure.

Does this mean our theory is useless?  Not at all: our theory not only
predicts the power laws, it also predicts the shape of the curves and
the way they cut off.  In particular, the avalanche size distribution
is predicted to have the following form as $r= (R- R_c)/R \to 0$:
$$D_{int}(S, R) \sim S^{-(\tau+\sigma\beta\delta)} 
			{\cal D}_{int}(s^\sigma r).\tag{curlyD}$$
At the critical point $r=0$: so long as ${\cal D}(0)\ne 0$, the distribution
is a pure power law with power $\tau+\sigma\beta\delta$.  Near the
critical point, it starts to deviate from a pure power law 
when the argument of ${\cal D}$ becomes near one --- that is,
when the avalanche size $s \sim r^{-1/\sigma}$.  Usually, the scaling
function ${\cal D}$ is constant for small arguments, and dies away
exponentially for large arguments.  Again, we emphasize that
the whole function ${\cal D}(x)$ is a universal property just as 
the critical exponents are.

\midinsert
\centerline{
\epsfxsize=2.2truein
\epsffile{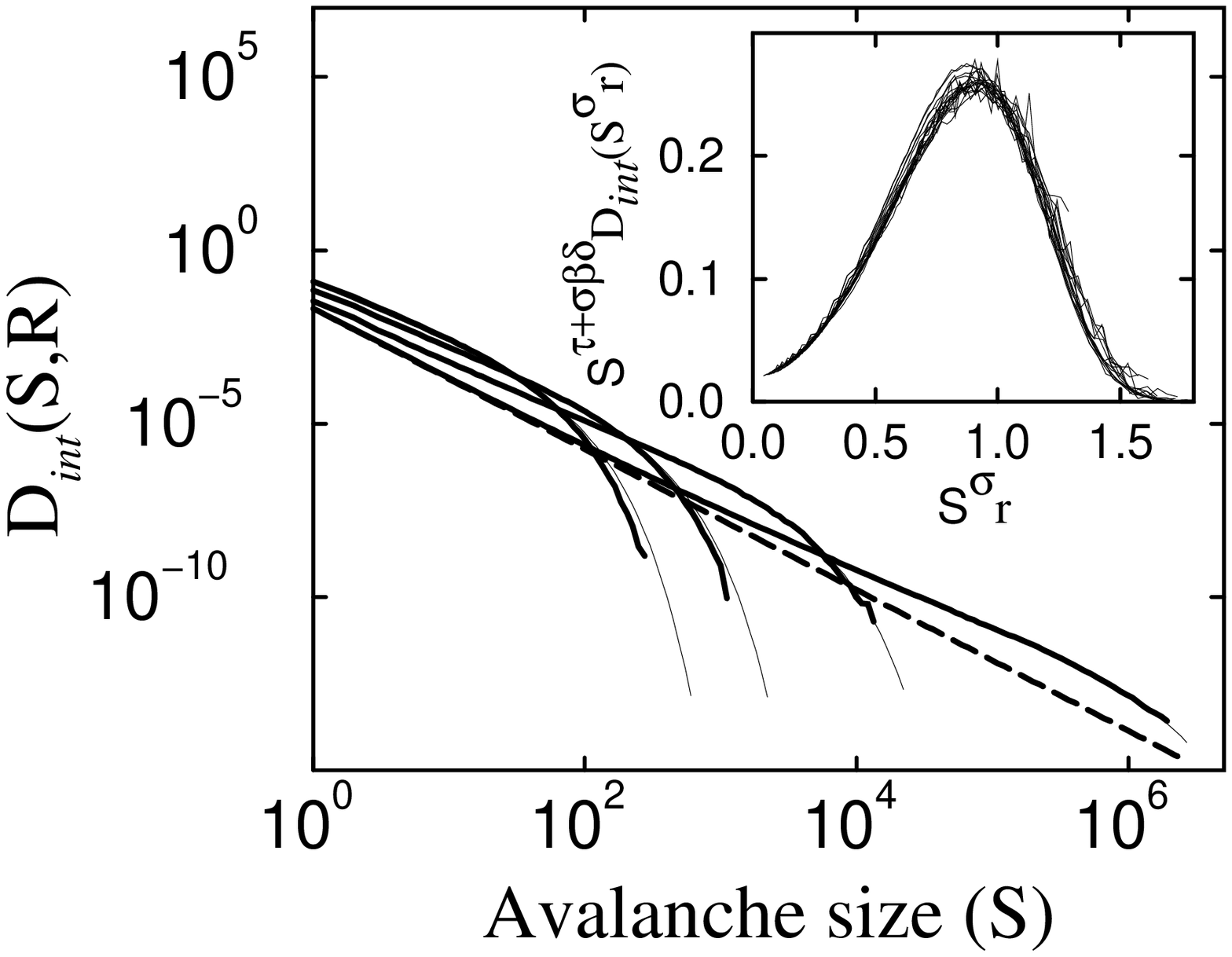}
\epsfxsize=2.2truein
\epsffile{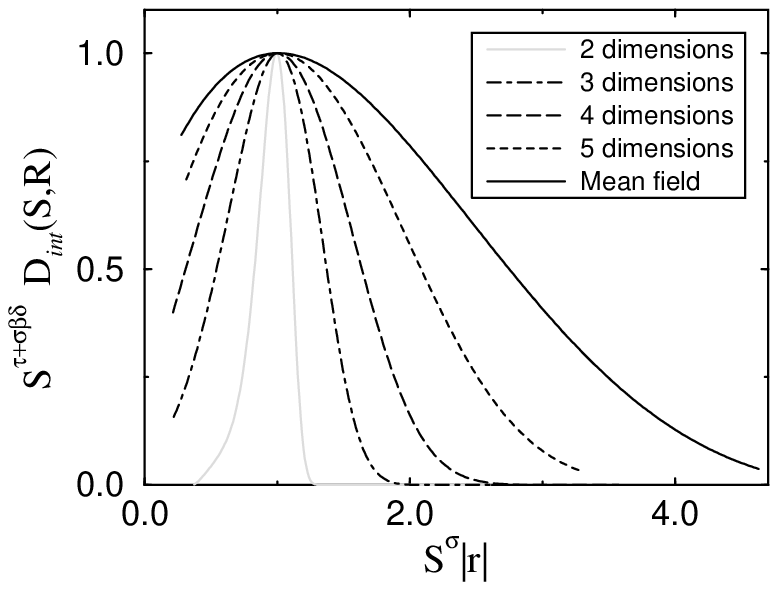}
}
\caption{Left, main:~The avalanche size distribution for our model in three
dimensions, for disorders $R = 4$, 3.2, 2.6, and 2.25.  The
smooth curves going through the data are the scaling predictions
of the theory.  The dashed line
shows the expected behavior at the critical disorder $R_c\sim 2.16$.
Left, inset:~The scaling collapses of these curves.  The reason the
slope of the avalanche size distribution converges so slowly is the
large bump in this curve: it grows by an order of magnitude from its
value at zero to the peak value.
Right:~The universal scaling curves in different dimensions.  The
explanation for the large bump in three dimensions is that the scaling
curve ${\cal D}(0)=0$ in two dimensions.  We believe
the exponent for the decay of the avalanche size distribution in two
dimensions is shifted by one, because of this zero\ref{perkovic}.
}
\endinsert

We can make a plot of the scaling function, by taking our data and
rescaling it:
$${\cal D}_{int}(s^\sigma (R-R_c)/R) \sim
	S^{\tau+\sigma\beta\delta} D_{int}(S,R). \tag{collapse}$$
That is, if you plot $S^{\tau+\sigma\beta\delta} D_{int}(S,R)$ versus
$s^\sigma (R-R_c)/R$, data taken at different R will all collapse onto the
same curve.  This is what Widom discovered in the early days of 
critical phenomena.  The data collapse is shown in the inset to
the left side of figure 5.  The theory tells us further that any other
system governed by the same universality class (one hopes someday
a real experiment) will also rescale onto this same curve.

Now we can understand why our pure power-law is so elusive.  The scaling
curve almost vanishes at $S^\sigma r = 0$: it rises by about an order of
magnitude before dying exponentially.  Indeed, each of the sets of data
shows a bulge of about a factor of ten above the pure power law.  (Why
is the bulge so big?  The right-hand side of figure~5 is our
explanation: it's because we're so close to two dimensions, where 
${\cal D}_{int}(0)=0$.) We can work backward from the scaling curve and make
predictions for the avalanche size distribution for each of the
different disorders: the smooth curves in figure~5 are predictions of
the scaling theory. The power law still isn't useful at $R=2.25$, where
$r=0.04$, but the complete Widom scaling prediction is quite successful
all the way out to $R=4$, almost twice the critical value.

\section{CONCLUSION}

So, we have an understanding of why the noise pulses in magnets can span
such a range of scales: they are near a critical point where the hysteresis
loop develops a jump.  We have a scaling description of the behavior near
the critical point.  We have a rough explanation of the experimental 
observations.  We also think we understand why the measurements
might fluctuate so far from our predictions: instead of varying a parameter 
and doing a scaling collapse, the experiments only measure an effective
power-law.  We recommend trying to get closer (or farther) from the critical
point.

Why do the experiments see power laws?  That is, why are the samples
they use near the critical point?  Unlike more traditional phase
transitions, our model has a large {\it critical range}: 4\% away from the
critical point we have six decades of scaling, and a factor of two away
we still have two decades.  Perhaps there are mechanisms which tune
the system precisely to the critical point, but it seems likely that the
experimentalists could just be lucky and pick their sample inside this
large range.

Finally, why is our critical range so large?  This technically doesn't
have a clean answer: the size of the critical range isn't a universal
property!  Universal questions aren't the only important ones, of course.
I think there are three contributing factors.  (1)~The critical exponent
$\nu = 1.42$ in our model, where in the three-dimensional Ising model
it is 0.63.  That means that getting twice as close to $R_c$ makes
the length spanned by an avalanche grow by a factor of 2.7, where getting
twice as close to $T_c$ for the Ising model makes the correlation length
grow only by a factor 1.55.  (2)~The size $S$ of an avalanche is more like
a volume than a length.  Six decades of scaling in $S$ should be thought
of as roughly two decades in length scale.  (Actually, since the avalanches
aren't space filling in three dimensions, the volume 
$S \sim \xi^{1/\sigma \nu} \sim \xi^{2.6}$, so six decades in size $S$
gives 2.3 decades in the length scale $\xi$.) 
(3)~Three dimensions is close to two dimensions.
As you can see from the right side of figure five, the behavior in three
dimensions is far removed from the mean-field behavior of the model
in six and higher dimensions.  The fluctuations are extremely important;
in two dimensions, we believe that they almost completely dominate, perhaps
even preventing an infinite avalanche from ever occurring!  If $R_c=0$
in two dimensions, then the critical regime must span to $\infty \times R_c$;
no wonder in three dimensions that it spans to $2 \times R_c$.

There are still many important unsolved problems here, but it's clear
that the traditional methods of critical phenomena --- Widom scaling,
the renormalization group, and the $\epsilon$ expansion --- have
been remarkably useful.

\ack{We acknowledge the support of DOE Grant \#DE-FG02-88-ER45364 and NSF
Grant \#DMR-9118065.  This work was partly conducted at
the Cornell National Supercomputer Facility, funded in part by NSF,
by NY State, and by IBM.  
Further pedagogical information (along with your own copy of the
simulation, in Java) is available at \hfil\break
\centerline{http://www.lassp.cornell.edu/sethna/hysteresis/hysteresis.html.}
}

\references

\refis{prl1}
J. P. Sethna, K. A. Dahmen, S. Kartha, J. A. Krumhansl,
	B. W. Roberts, and J. D. Shore, {\it Phys. Rev. Lett.},
	{\bf 70}, 3347 (1993).
J. P. Sethna, K. Dahmen, S. Kartha, J. A. Krumhansl, O. Perkovi\'{c},
	B. W. Roberts, and J. D. Shore (reply), {\it Phys. Rev. Lett.}
	{\bf 72}, 947 (1994).

\refis{prl2}
K. A. Dahmen and J. P. Sethna, {\it Phys. Rev. Lett.} {\bf 71},
	3222 (1993).

\refis{prl3}
O. Perkovi\'{c}, K. A. Dahmen, and J. P. Sethna, {\it Phys. Rev.  Lett.}
	{\bf 75}, 4528 (1995).

\refis{prb}
K. A. Dahmen and J. P. Sethna, {\it Phys. Rev. B} {\bf 53}, 14872 (1996).

\refis{perkovic} O. Perkovi\'{c}, K. A. Dahmen, and J. P. Sethna,
Condensed-Matter Archive preprint \#9609072.

\refis{new}
Olga Perkovi\'{c} and James P. Sethna, {\it J.  Appl. Phys.} {\bf 81}, 1590
(1997).
Deepak Dhar, Prabodh Shukla, James P. Sethna, (submitted to {\it J. Phys. A}).

\refis{experiments}
Barkhausen noise in different magnetic
materials: Fe, alumel, metglass [P. J. Cote and L. V. Meisel,
{\it Phys. Rev. Lett.} 67, 1334 (1991);
L. V. Meisel and P. J. Cote, {\it Phys. Rev. B} 46, 10822 (1992)],
NiS [K. Stierstadt and W. Boeckh, {\it Z. Physik} {\bf 186}, 154 (1965)],
SiFe [G. Bertotti, G. Durin, and A. Magni
{\it J. Appl. Phys.} {\bf 75}, 5490 (1994),
H. Bittel {\it IEEE Trans. Magn.} {\bf 5}, 359 (1969),
U. Lieneweg
{\it IEEE Trans. Magn.} {\bf 10}, 118 (1974)],
81$\%$NiFe [U. Lieneweg and W. Grosse-Nobis
{\it Intern. J. Magnetism} {\bf 3}, 11 (1972)],
AlSiFe [G. Bertotti, F. Fiorillo, and A. Montorsi
{\it J. Appl. Phys.} {\bf 67}, 5574 (1990)],
and FeNiCo [J.S. Urbach, R.C. Madison, and J.T. Markert, 
{\it Phys.\ Rev.\ Lett.} {\bf 75}, 276 (1995)].
The sample shapes were mostly wires. The quoted exponents were
experimentally obtained
from the pulse-area distribution in a small
bin of the magnetic field $H$ (exponent $\tau$), the pulse-area
distribution integrated over the entire hysteresis loop $(\tau +
\sigma \beta \delta)$, the distribution of pulse durations in a
small
bin of $H$ ($(\tau-1)/\sigma\nu z + 1$), the distribution of
pulse durations integrated over the loop ($(\tau + \sigma\beta\delta
-1)/\sigma \nu z + 1$), the power spectrum of the pulses
in a small bin of $H$ ($(3-\tau)/\sigma\nu z$), the power spectrum
of the pulses integrated over the hysteresis loop
($(3-(\tau+\sigma\beta\delta))/\sigma\nu z$), and the distribution of
pulse energies in a small bin of $H$ ($(\tau-1)/(2-\sigma\nu z) + 1$).
Notice that these experiments are mostly done in geometries which minimize
the effects of demagnetization fields.

\refis{epsilon5} H. Kleinert, J. Neu, V. Schulte-Frohlinde,
K.G. Chetyrkin, and S. A. Larin ({\it Phys. Lett. B} {\bf 272}, 39
(1991) and erratum in {\it Phys. Lett. B} {\bf 319}, 545 (1993));
A.~A.~Vladimirov, D.~I.~Kazakov, and O.~V.~Tarasov,
{\it Sov. Phys. JETP} {\bf 50} (3), 521 (1979) and references therein;
J.C. LeGuillou and J. Zinn-Justin {\it Phys.  Rev. B} {\bf 21}, 3976 (1980),
{\it J. Physique Lett.} {\bf 46}, L137 (1985), {\it J. Physique} {\bf 48}, 
19 (1987);  J. Zinn-Justin ``Quantum Field Theory and Critical Phenomena'', 
2nd edition, Clarendon Press, Oxford (1993)); 
G.~Parisi, lectures given at the 1982 Les
Houches summer school XXXIX ``{\it Recent advances in field theory and
statistical mechanics}'' (North Holland), and references therein.

\refis{depinning} Mark Robbins introduced this model, (allowing only 
connected spins to flip), in the study of the depinning transition
({\it e.g.} in fluid invasion): Hong Ji and Mark O. Robbins, {\sl Phys. Rev. B}
{\bf 46}, 14519 (1992) and references therein.

\refis{epsilon}  The $\epsilon$-expansion has a vast literature:
recent works to study depinning transitions that inspired us include
O.~Narayan and D.~S. Fisher, Phys. Rev. Lett. {\bf 68}, 3615 (1992), 
Phys. Rev. B {\bf 46}, 11520 (1992), and Phys. Rev. B {\bf 48}, 7030 (1993);
T. Nattermann, S. Stepanow, L. H. Tang and H. Leschhorn, J. Phys. II France 
{\bf 2}, 1483 (1992)

\refis{simulations} Other models have been studied in a similar context:
see C. M.~Coram, A.~E.~Jacobs, N.~Heinig, and K. B. Winterbon, {\sl Phys. Rev.
B} {\bf 40}, 6992 (1989); E. Vives and A. Planes, {\sl Phys. Rev. B} {\bf 50},
3839 (1994); E. Vives, J. Goicoechea, J. Ort\'in, and A. Planes, {\sl Phys. 
Rev. E} {\bf 52}, R5 (1995).

\endreferences

\bye